\documentclass[a4paper,10pt,twoside]{cpc-hepnp}
\usepackage{CJK,upgreek,fancyhdr}
\usepackage{multicol}
\usepackage{graphicx}
\usepackage{booktabs}
\usepackage{amssymb,bm,mathrsfs,bbm,amscd}
\usepackage[tbtags]{amsmath}
\usepackage{lastpage}

\usepackage{color}
\usepackage{slashed}
\usepackage{tabularx}

\begin{document}
\begin{CJK*}{GB}{gbsn}

\fancyhead[c]{\small Chinese Physics C~~~Vol. xx, No. x (201x) xxxxxx}
\fancyfoot[C]{\small 010201-\thepage}
\footnotetext[0]{}

\title{Deuteron electromagnetic form factors with the light-front approach\thanks{Supported by National Natural Science
Foundation of China (No.~10975146, and No.~11475192). The fund provided by the Sino-German CRC 110 ``Symmetries and the Emergence of Structure in QCD" project is also appreciated. YBD thanks FAPESP grant 2011/11973-4 for funding his visit to ICTP-SAIFR. }}

\author{%
      Bao-dong Sun  $^{1;1)}$\email{sunbd@ihep.ac.cn}%
\quad Yu-bing Dong  $^{1,2;2)}$\email{dongyb@ihep.ac.cn}%
}
\maketitle

\address{%
$^1$ 
Institute of High Energy Physics, Beijing 100049, P. R. China\\
$^2$
Theoretical Physics Center for Science Facilities (TPCSF), CAS,
Beijing 100049, P. R. China
}
\begin{center}
\date{\today}
\end{center}

\begin{abstract}

The electromagnetic form factors and low-energy observables of the deuteron are studied with the help of the light-front approach, where the deuteron is regarded as a weakly bound state of a proton and a neutron.  Both the $S$ and $D$ wave interacting vertexes among the deuteron, proton, and neutron are taken into account. Moreover, the regularization functions are also introduced. In our calculations, the vertex and the regularization functions are employed to simulate the momentum distribution inside the deuteron. Our numerical results show that the light-front approach can roughly reproduce the deuteron electromagnetic form factors, like charge $G_0$, magnetic $G_1$, and quadrupole $G_2$, in the low $Q^2$ region. The important effect of the $D$ wave vertex on $G_2$ is also addressed.

\end{abstract}



\footnotetext[0]{\hspace*{-3mm}\raisebox{0.3ex}{$\scriptstyle\copyright$}2013
Chinese Physical Society and the Institute of High Energy Physics
of the Chinese Academy of Sciences and the Institute
of Modern Physics of the Chinese Academy of Sciences and IOP Publishing Ltd}%

\begin{multicols}{2}

\section{Introduction}

The study of deuteron properties, like its mass, binding energy, radius, and electromagnetic form factors, has been of great interest for many years, since it can encode the nature of nuclear effects and the nucleon--nucleon interaction. The deuteron, with spin-1, is one of the simplest nuclei, and it is usually regarded as a loosely bound state of a proton and a neutron. This feature also makes the deuteron a widely used substitute for a neutron target or neutron beams. Moreover, as the only two-nucleon bound state, the study of the deuteron is a good starting point to understand multi-nucleon systems.

Under the one-photon-exchange (OPE) approximation, it is usually believed~\citep {Gourdin1974} that the deuteron electromagnetic current can be expressed by the sum of the two triangular diagrams in Fig.~\ref{figure:PhotonDeuteronPN}, and the deuteron ground state is approximately spherical symmetric (see the discussion in Ref.~\cite{Garcona2001}, for example) with a small mixture of $D$ wave. According to the work in relativistic quantum mechanics by Chung, Coester and Keiser~\citep{Chung1988}, the electromagnetic current matrix, for a particle with intrinsic spin $S\geqslant1$,  only has $(S+1)^2-1$ or $(2S+1)(2S+3)/4-1$ independent components respectively for the integer or half-odd spin due to its Hermitian and its rotational invariance properties. In the deuteron case, with $S=1$, there are, therefore, three independent current matrix elements corresponding to the three conventional form factors, charge $G_0$, magnetic $G_1$, and quadrupole $G_2$.

There are many works in the literature which investigate the deuteron properties with the help of  non-relativistic potential models, effective Lagrangian approaches, relativistic frameworks, and many others~\cite{Gross2003,Chung1988,Sick2001,Gilman2002,Arnold1980,Mathiot1989,Arenhovel2000,Buchmann1989,Tjon1989,Kaplan1999,Ramalho2008,Karmanov1994,Allen2001,Carbonell1999,Phillips1998,Chen1998,Ivanov1995,Lev2000,Dong2008,Garcona2001,
Carbonell1998,Cooke200209}. Two relativistic approaches, the Bethe-Salpeter formalism and the light-front approach, have been widely employed to describe  bound state problems like the deuteron~\citep{Carbonell1998,Cooke200209}. The light-front approach has been successfully applied for the pion form factors ~\citep{Lepage1980, Frederico2009, Frederico2010, Gutsche2015}, for the $K$ and $\rho$ meson form factors~\citep{Pereira2007,Machleidt2001,Frederico1997}, and for the distribution amplitudes and decay constants of $\pi$, $K$ and $\rho$ et al~\citep{Choi2007}.

In the study of the $\rho$ meson (another system with $S=1$) properties by Ref.~\citep{Frederico1997}, the three conventional form factors $G_0$, $G_1$ and $G_2$ are extracted from the matrix elements of  the front-form electromagnetic current $J^+(=J^0+J^3)$. In the covariant light-front formalism, the four-vectors are given on the hypersurface specified by the invariant condition $n\cdot x=0$, where $x=(t,\vec x)$ and $n$ is an arbitrary light-like four-vector, that is, $n^2=0$. However, in the usual light-front formulation, $n=(1,0,0,-1)$ is always taken and the light-front hypersurface is given by $x^+=t+z=0$. Due to the approximation in the actual calculations, the final results may depend on the particular choices of the orientation of the light-front plan~\citep{Carbonell1998,Cooke200209}. Since the stability group no longer contains the rotation generators around the $x$ and $y$ axes, the rotation symmetry around $x$ and $y$ axes, which respectively correspond to the angular conditions $J^+_{yy}=J^+_{zz}$ and $J^+_{xx}=J^+_{zz}$ (where the subscripts are the polarizations in the instant form spin basis) may both be broken~\citep{Frederico1997}. Therefore, there are several different approaches to obtain the form factors in the light-front framework~\citep{Frankfurt1989,Chung1988,Arnold1980,Frankfurt1993}.

To find which prescription is more suitable for the study of the $\rho$ meson properties, Melo and Frederico~\citep{Frederico1997} compared the calculation of the non-covariant light-front approach with the covariant case. In their work, the covariant calculation is done by integrating the  $k^0$ component of the loop momentum analytically and the rest numerically. In the light-front calculation, the $k^-(=k^0-k^3)$ component is integrated analytically and $x(=k^+/p^+)$ (where $k^+=k^0+k^3$ and $p^+=p^0+p^3$) and ${\bf k}_{\perp}(=(k^1,k^2))$ are integrated numerically. Here, the Cauchy integral with respect to $k^-$ is carried out in the Breit frame with the $q^+$  component of momentum transfer vanishing. This selection leaves only one pole valid in the loop integration, corresponding to the forward propagator. The numerical results in Ref.~\citep{Frederico1997} show that different extraction prescriptions cause sizeable effects on the form factors and static properties of the $\rho$ meson.

In this work, we plan to apply  the light-front approach of Ref.~\citep{Frederico1997} to the  calculation of  the deuteron form factors. Here the extraction prescription, proposed by Frankfurt, Frederico, and Strikman~\citep{Frankfurt1993}, will be taken into account. In our calculation, we regard the deuteron as a weakly bound state of a proton and a neutron, and we do the numerical calculation in Minkowski space for the loop integral with the help of the light-front approach. Therefore, the present work is different from our previous calculations which were done in Euclidean space with a Gauss-type regularization~\citep{Dong2008,Liang2015}. In our calculation, the model-dependent parameters will be determined by fitting the form factors to the experimental data. Moreover, we employ the empirical parametrization forms ~\citep{Blunden2005} for the  $\gamma-p$ and $\gamma-n$ vertexes for our numerical calculations. For the $S$ wave spin structure of the vertex among
the deuteron, proton and neutron, we take the form proposed by Ref.~\citep{Frederico1997} for the $\rho$ meson.
In order to account for the $D$ wave spin structure, we refer to the work of Blankenbecler, Gloderber, and Halpern~\citep {Blankenbecler1959}.

This work is organized as follows. In Section 2, the framework of our calculations is briefly shown and the four prescriptions for the extraction of the form factors, in the light front approach, are explicitly discussed. In Section 3, the light-front current $J^+$ is constructed from the one-loop diagram shown in Fig.~\ref{figure:PhotonDeuteron}. To get a finite loop integral in Minkowski space and to get a better simulation of momentum distributions of the proton and neutron inside the deuteron, the regularization functions are also employed. In Section 4, a set of parameters is given by fitting the obtained form factors to the experimental data, and the numerical results for the deuteron low-energy properties are also displayed. Moreover, a detailed discussion about the relations between the $S$ and $D$ vertex structures  and their effects on the form factors are also displayed  in this section. Finally, a short summary is given in the last section.

\begin{center}
\includegraphics[width=8cm]{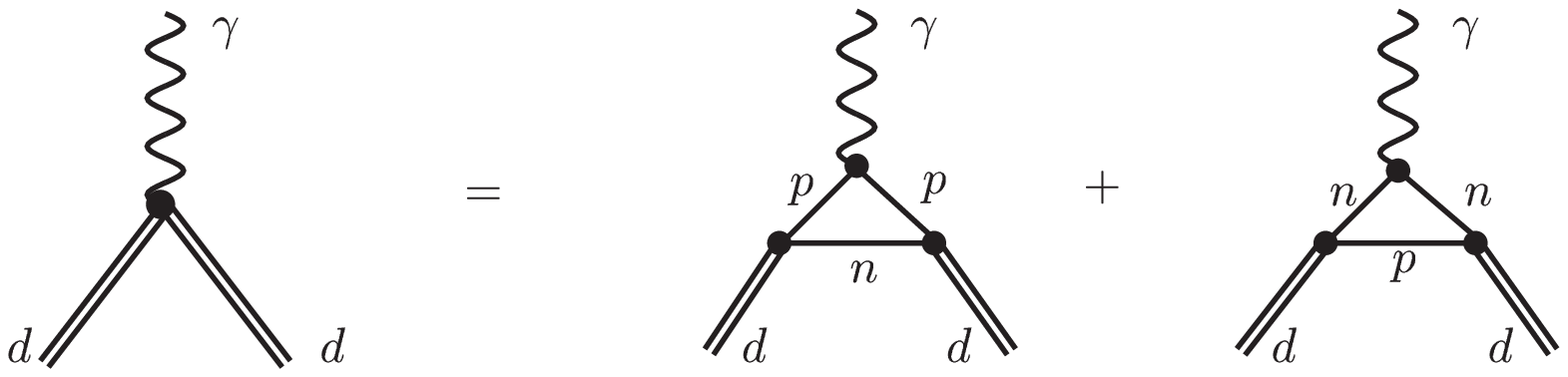}
\figcaption{\label{figure:PhotonDeuteronPN} {\small  The deuteron electromagnetic vertex.} }
\end{center}

\section{Theoretical framework}
In the OPE (Born) approximation, by neglecting the electron mass $m_e\sim0$, one can get the conventional form of the unpolarized e-d elastic scattering differential cross section as~\citep{Coester1975}
\begin{eqnarray}
\frac{d\sigma}{d\Omega}&=&{\sigma}_M \left( A(Q^2) + B(Q^2)tan^2\frac{\theta}{2} \right),
\end{eqnarray}
with $\theta$ being the scattering angle and the two structure functions
\begin{eqnarray}
A(Q^2)&=&G^2_0(Q^2) + \frac{8}{9}\eta^2 G^2_2(Q^2)  + \frac{2}{3}{\eta}G^2_1(Q^2) \ ,
\nonumber \\
B(Q^2)&=&\frac{4}{3}\eta(1+\eta)G^2_1(Q^2),
\label{eq:AandB}
\end{eqnarray}
where $\eta=Q^2/4m^2_D$  with $m_D$ being the deuteron mass, $Q^2\equiv-q^2$ with $q$ being the momentum transfer, and ${\sigma}_M$ is the Mott cross section. To extract the three form factors $G_0$, $G_1$ and $G_2$, except for the two structure functions $A(Q^2)$ and $B(Q^2)$, one usually needs another observable, like the tensor polarization $t_{20}$, to determine the three form factors.

The electron-deuteron elastic scattering process can be treated as the interaction between electron and deuteron electromagnetic currents. The corresponding amplitude can be written as
\begin{eqnarray}
\it \mathcal{M}_{jk}&=&\frac{e^2}{Q^2}\overline{u}(l')\gamma_\mu u(l) J^{\mu}_{jk}(p_f,p_i)\ , \
\end{eqnarray}
where the subscripts $(j, k)=(x,y,z)$ stand for the indices of the final and initial deuteron polarizations and $p_f$ and $p_i$ are the deuteron final and initial momenta. We know that there are three independent electromagnetic current elements of the spin-1 particle, then the deuteron current matrix element as $J_{jk}^\mu(p_f,p_i)={\epsilon}^{'*\alpha}_j S^\mu_{\alpha\beta} {\epsilon}^{\beta}_k$ can be factorized by the three form factors, where $\epsilon^{'\alpha}_j$ and  $\epsilon^\beta_k$ are respectively the final and initial polarization vectors, which will be defined later. $S^\mu_{\alpha\beta}$ can be written as the general form~\citep{Frederico1997}
\begin{eqnarray}
\label{eq:curr1}
S_{\alpha \beta}^{\mu}&=&\Big [F_1(Q^2)g_{\alpha \beta}-F_2(Q^2)
\frac{Q_{\alpha}Q_{\beta}}{2 m_D^2}\Big ] P^\mu \nonumber\\[1mm]
&&
-F_3(Q^2)(Q_\alpha g_\beta^\mu-
Q_\beta g_\alpha^\mu) \ ,
\end{eqnarray}
where $P^\mu$ is the sum of $p^\mu_i$ and $p^\mu_f$.
The charge monopole $G_0$, magnetic dipole $G_1$ and charge quadrupole $G_2$ form factors in Eq.~(\ref{eq:AandB}) relate to the form factors of $F_1$, $F_2$ and $F_3$ by~\citep{Garcona2001}
\begin{eqnarray}
G_0(Q^2)&=&F_1(Q^2) + \frac{2}{3}{\eta} G_2(Q^2) \ , \ \\
G_1(Q^2)&=&F_2(Q^2) \ , \ \\
G_2(Q^2)&=&\frac{3\sqrt{2}}{4\eta}(F_1(Q^2) - F_2(Q^2) + (1+\eta)F_3(Q^2))\ , \
\nonumber\\
\end{eqnarray}
and they are normalized to
\begin{eqnarray}
G_0(0)=1\ , \ G_1(0)=\frac{m_D}{m}\mu_d \ , \ \frac{G_2(Q^2)}{\frac{2\sqrt{2}}{3}\eta}\bigg\arrowvert_{Q^2 = 0}=m_D^2 Q_d \ , \
\end{eqnarray}
where $m$ is the nucleon mass, $\mu_d=0.857\mu_N$ is the deuteron magnetic moment in units of the nuclear magneton $\mu_N$, and $Q_d=0.286 $fm$^2$ is the deuteron quadrupole moment.

In the Breit frame with the instant form spin basis, the momentum transfer is chosen to be in the positive $x$ direction, $q^{\mu}=p^\mu_f-p^\mu_i=(0,q_x,0,0)$, with $p^\mu_i=(p^0,-q_x/2,0,0)$ and $p^\mu_f=(p^0,q_x/2,0,0)$, where $p^0=m_D\sqrt{1+\eta}$. Thus, the corresponding angular condition of $J^+_{yy}=J^+_{zz}$ breaks
down due to this specified reference frame as discussed above. The deuteron Cartesian polarization four-vectors are
\begin{eqnarray}
\epsilon^\mu_x&=&(-\sqrt{\eta},\sqrt{1+\eta},0,0) \ , \
\epsilon^\mu_y =(0,0,1,0) \ , \
\nonumber \\
\epsilon^\mu_z &=& (0,0,0,1) \ , \
\label{eq:intpv}
\end{eqnarray}
for the initial state, and
\begin{eqnarray}
\epsilon'^\mu_x=(\sqrt{\eta},\sqrt{1+\eta},0,0) \ , \
\epsilon'^\mu_y =\epsilon^\mu_y \ , \
\epsilon'^\mu_z = \epsilon^\mu_z \ , \
\label{eq:finpv}
\end{eqnarray}
for the final state.

To get the conventional electromagnetic form factors, one needs to transfer the Cartesian polarization four-vectors to the spherical spin basis:
\begin{eqnarray}
\epsilon_{\pm} ^{(')}=\mp \frac{\epsilon_x^{(')}\pm \epsilon_y^{(')}}{\sqrt{2}}~\ ,~~~~~
\epsilon_{0}^{(')}=\epsilon_{z}^{(')} ,
\end{eqnarray}
for the initial (or final)  state.
With this spin basis set, the $J^+$ component of the electromagnetic current has the form~\citep{Frankfurt1993}
\begin{eqnarray}
J^{+}   =  \frac{1}{2}  \left( \begin{array}{ccc}
J_{xx}^{+}+J^+_{yy}&  -\sqrt{2} J^+_{zx}    &  J_{yy}^{+}-J^+_{xx} \\
\sqrt{2} J^+_{zx}  &  2 J^+_{zz}  &  -\sqrt{2} J^+_{zx}          \\
J^+_{yy}-J^+_{xx}  &     \sqrt{2} J^+_{zx}   &  J_{xx}^{+}+J^+_{yy}  \\
\end{array} \right) \ ,
\label{eq:inst}
\end{eqnarray}
where the order of the projection is $(+,0,-)$.

It is convenient to extract the three form factors from the $I^+$ component of the light-front form electromagnetic current. In the front-form spin basis, the plus components of the electromagnetic current matrix $I^+_{\lambda\lambda'}$ have the form
\begin{eqnarray}
I^+ =  \left( \begin{array}{ccc}
I_{11}^{+}  & I_{10}^{+}  & I_{1-1}^{+} \\
  - I_{10}^{+}  & I_{00}^{+}  &  I_{10}^{+}         \\
I_{1-1}^{+}  &   - I_{10}^{+}   & I_{11}^{+}  \\
\end{array}  \right) \ ,
\label{eq:ifront}
\end{eqnarray}
where the subscripts $(\lambda, \lambda')=(\pm1, 0)$ label the different polarizations in the front-form spin basis.
The unitary transformation between the instant-form spin basis and the front-form spin basis is the Melosh rotation~\citep{Melosh1974,Frederico1991}. With the help of the Melosh rotation, $I^+_{\lambda\lambda'}$ can be expressed by $J^+_{jk}$  as~\citep{Frederico1997}
\begin{eqnarray}
I^{+}_{11}&=&\frac{J^{+}_{xx}+(1+\eta) J^{+}_{yy}-
\eta J^{+}_{zz}+2 \sqrt{ \eta} J^{+}_{zx}}{2 (1+\eta)}
\nonumber \\
I^{+}_{10}&=&\frac{\sqrt{2 \eta} J^{+}_{xx}+\sqrt{2 \eta} J^{+}_{zz}
+\sqrt{2} (\eta-1) J^{+}_{zx}}{2(1+\eta)}
\nonumber \\
I^{+}_{1-1}&=&\frac{-J^{+}_{xx}+(1+\eta) J^{+}_{yy}+
\eta J^{+}_{zz}-2 \sqrt{\eta} J^{+}_{zx}}{2 (1+\eta)}
\nonumber \\
I^{+}_{00}&=&\frac{-\eta J^{+}_{xx}+J^{+}_{zz}+2 \sqrt{\eta} J^{+}_{zx}}
{(1+\eta)} \ .
\label{eq:ifront1}
\end{eqnarray}

Since the rotational invariance condition breaks down, there several different ways to extract the form factors~\citep{Frederico1997}.  For example, one may consider some of components as
 ``good'' ones and keep them, and neglect the ``worst" one. More details can be found in Refs.~\citep{Grach1984, Cooke200209, Chung1988, Brodsky1992, Frankfurt1993}.

In the work of Grach and Kondratyuk (GK)~\citep{Grach1984,Frankfurt1989}, they chose the  ``worst'' component as $I^{+}_{00}$ and then got
\begin{eqnarray}
G_0^{GK}&=&\frac{1}{3}[(3-2 \eta) I^{+}_{11}+
2 \sqrt{2 \eta} I^{+}_{10} +  I^{+}_{1-1}]
\nonumber\\
&=&\frac{1}{3}[J_{xx}^{+}+ 2 J_{yy}^{+}-\eta  J_{yy}^{+}
+ \eta  J_{zz}^{+}]
\nonumber \\
G_1^{GK}&=&2 [I^{+}_{11}-\frac{1}{ \sqrt{2 \eta}} I^{+}_{10}]
\nonumber\\
&=&J_{yy}^{+} -  J_{zz}^{+}+\frac{J_{zx}^{+}}{\sqrt{\eta}}
\nonumber \\
G_2^{GK}&=&\frac{2 \sqrt{2}}{3}[- \eta I^{+}_{11}+
\sqrt{2 \eta} I^{+}_{10} -  I^{+}_{1-1}]
\nonumber\\
&=&\frac{\sqrt{2}}{3}[J_{xx}^{+}+J_{yy}^{+} (-1-\eta)
+\eta  J_{zz}^{+}] \ .
\label{eq:Frankfurt1989}
\end{eqnarray}
\noindent
In the work of Chung, Coester, Keister and Polizou (CCKP)~\citep{Chung1988}, on the other hand, they kept all the four components $I^{+}_{11}$, $I^{+}_{00}$, $I^{+}_{10}$ and $I^{+}_{1-1}$ and obtained
\begin{eqnarray}
G_0^{CCKP}&=&\frac{1}{3 (1+\eta)}
[(\frac{3}{2}-\eta) (I^{+}_{11}+I^{+}_{00})
+5 \sqrt{2 \eta} I^{+}_{10}
\nonumber\\
&&
+(2 \eta - \frac{1}{2}) I^{+}_{1-1}]
\nonumber \\
&=&\frac{1}{6}[2 J_{xx}^{+} + J_{yy}^{+} + 3 J_{zz}^{+}]
\nonumber \\
G_1^{CCKP}&=&\frac{1}{(1+\eta)}[I^{+}_{11}+I^{+}_{00}
-I^{+}_{1-1}  - \frac{2 (1-\eta)}{\sqrt{2\eta}} I^{+}_{10}]
\nonumber\\
&=&\frac{J_{zx}^{+}}{\sqrt{\eta}}
\nonumber \\
G_2^{CCKP}&=&\frac{\sqrt{2}}{3 (1+\eta)}[- \eta I^{+}_{11}
-\eta I^{+}_{00}+2 \sqrt{2 \eta} I^{+}_{10}
\nonumber\\
&&
- (\eta + 2) I^{+}_{1-1}]
\nonumber\\
&=&\frac{\sqrt{2}}{3} [J_{xx}^{+}-J_{yy}^{+}]\ .
\label{eq:Chung1988}
\end{eqnarray}
\noindent
Moreover, in the work of Brodsky and Hiller (BH)~\citep{Brodsky1992,Arnold1980}, the component $I^{+}_{11}$ was avoided, and they gave
\begin{eqnarray}
G_0^{BH}&=&\frac{1}{3(1+\eta)}[(3-2 \eta) I^{+}_{00}+8 \sqrt{2 \eta} I^{+}_{10}
\nonumber\\
&&
+2 (2 \eta -1) I^{+}_{1-1}]
\nonumber \\
&=&\frac{1}{3 (1+2 \eta)}[J_{xx}^{+} (1+2 \eta)+ J_{yy}^{+}(2 \eta-1)
\nonumber\\
&&
+ J_{zz}^{+}(3+2 \eta)]
\nonumber \\
G_1^{BH}&=&\frac{2}{(1+2 \eta)}[I^{+}_{00}
-I^{+}_{1-1}+\frac{(2 \eta -1)}{\sqrt{2 \eta}} I^{+}_{10}]
\nonumber \\
&=&\frac{1}{(1+2 \eta)}[\frac{J_{zx}^{+}}{\sqrt{\eta}}
 (1+2 \eta)- J_{yy}^{+} +  J_{zz}^{+}]
\nonumber \\
G_2^{BH}&=&\frac{2 \sqrt{2}}{3 (1+ 2 \eta)}[\sqrt{2 \eta} I^{+}_{10}
-\eta I^{+}_{00} -( \eta+1) I^{+}_{1-1}]
\nonumber \\
&=&\frac{ \sqrt{2}}{3 (1+2 \eta)}[J_{xx}^{+} (1+2 \eta)- J_{yy}^{+}(1+ \eta)
\nonumber\\
&&
- \eta J_{zz}^{+}]\ . \
\label{Brodsky1992}
\end{eqnarray}
\noindent
In this work, we will apply the approach of Frankfurt, Frederico and Strikman's prescription (FFS)~\citep{Frankfurt1993} to
the extraction of the deuteron form factors. It coincides with the CCKP prescription for $G_1$ and $G_2$ but differs for $G_0$,
\begin{eqnarray}
G_0^{FFS}&=&\frac{1}{3 (1+\eta)}[(2 \eta+3) I^{+}_{11}+2 \sqrt{2 \eta}I^{+}_{10} -\eta I^{+}_{00}
\nonumber\\
&&
+(2 \eta +1) I^{+}_{1-1}]
\nonumber\\
&=&\frac{1}{3}[J^{+}_{xx}+2 J^{+}_{yy}]
\nonumber \\
G_1^{FFS}&=&G_1^{CCKP}
\nonumber \\
G_2^{FFS}&=&G_2^{CCKP}
\ .
\label{eq:Frankfurt1993}
\end{eqnarray}
\noindent
It is straightforward to verify that those four prescriptions would be equivalent if the angular condition $J^{+}_{yy}= J^{+}_{zz}$ keeps valid. The differences between the four prescriptions will be analyzed later.

In addition, we know that the mean square charge radius $<r^2>$, magnetic moment $\mu_d$, and quadrupole moment $Q_d$ have the following relations to the three form factors~\citep{Chung1988},
\begin{eqnarray}
<r^2>&=&\lim_{Q^2 \rightarrow 0} \frac{6 (G_0(Q^2)-1)}{Q^2} \ , \
\mu_d=\lim_{Q^2 \rightarrow 0}\frac{m}{m_D}G_1(Q^2)  \ , \
\nonumber \\
Q_d&=&\lim_{Q^2 \rightarrow 0} 3\sqrt{2}\frac{G_2(Q^2)}{Q^2} \ .
\label{eq:LowEO}
\end{eqnarray}
Therefore, those low-energy observables can also be determined if the form factors are calculated.

\section{The light-front current}

To perform  the numerical calculation for the electromagnetic matrix elements, we consider both the $S$ and $D$ wave
vertex functions for the d-pn interaction,
\begin{eqnarray}
{\Gamma}^{\mu}_{d-pn}&=&{\Gamma}^\mu_S+{\Gamma}^\mu_D\ , \
\label{eq:FullVertex}
\end{eqnarray}
where the $S$ wave vertex takes the form proposed in Ref.~\citep{Frederico1997},
\begin{eqnarray}
{\Gamma}^{\mu}_S\big (k,k-p_{i,f}  \big)&=&{\gamma}^{\mu}-\frac{m_D}{2}\frac{2k^{\mu}-p^{\mu}_{i,f}}{p_{i,f}\cdot k+m_Dm-i\epsilon}\ , \
\label{eq:SwaveVertex}
\end{eqnarray}
and the $D$ wave vertex has been explicitly studied by Blankenbecler, Goldberger, Halpern~\citep{Liang2015,Blankenbecler1959}, and can be expressed as
\begin{eqnarray}
{\Gamma}^{\mu}_D\big (k,k-p_{i,f}  \big)&=&\rho\bigg (
{\gamma}^{\mu}-\frac{3}{m\delta}\big (\slashed{k}- \frac{\slashed{p}_{i,f}}{2} \big ){\gamma}^{\mu}\big (\slashed{k}- \frac{\slashed{p}_{i,f}}{2}\big )\bigg )\ , \
\nonumber\\
\label{eq:DwaveVertex}
\end{eqnarray}
where  $\delta$ is the deuteron binding energy, and $\rho$ is a model-dependent parameter. $\rho$  is not an observable~\citep{Garcona2001}, and it connects to  the $D$ wave admixture in the nonrelativistic potential model.

In the practical calculation, we employ the following parametrization forms for the known
electromagnetic current of the nucleon
\begin{eqnarray}
{\Gamma}_{\gamma-NN}^\mu&=&F_1(Q^2){\gamma}^{\mu}+\frac{i{\sigma}^{\mu\nu}q_\nu}{2m}F_2(Q^2),
\end{eqnarray}
where $F_1(Q^2)$ and $F_2(Q^2)$ are the Dirac and Pauli form factors and they have been parameterized as
the sum of three or two monopoles, proposed by  Ref.~\citep{Blunden2005},
\begin{eqnarray}
F_{1,2}(Q^2)
&=& \sum_{i=1}^N { n_i \over d_i + q^2 }\ ,
\label{eq:3pole}
\end{eqnarray}
where $n_i$ and $d_i$ are parameters shown in Table~\ref{tab:param}. Note that for $F_2^n$, $N=2$.

\begin{table}
\end{table}
\begin{center}
\tabcaption{ \label{tab:param} {\small Parameters for $F_1$ and $F_2$ in
	Eq.~(\protect\ref{eq:3pole}) used in this work, with $n_i$,	$d_i$, and $Q^2$
in units of GeV$^2$.}}
\footnotesize

\begin{tabular}{c|rr|rr}				\hline
	& $F_1^p$\ \ \ \ \ & $F_2^p$\ \ \ \ \
	& $F_1^n$\ \ \ \ \ & $F_2^n$\ \ \ \ \		\\ \hline
\ $N$\	& 3\ \ \ \ \ & 3\ \ \ \ \ & 3\ \ \ \ \ & 2\ \ \ \ \  \\ \hline
\ $n_1$\ &\ 0.38676\ & 1.01650\  & 24.8109\   &\ 5.37640\  \\
\ $n_2$\ &\ 0.53222\ & --19.0246\ & --99.8420\ &\ --5.29920\  \\
\ $n_3$\ &\ --0.94491\ & 18.0371\ & 75.0544\ &\  ---\ \ \ \   \\
\ $d_1$\ &\ 3.29899\ & 0.40886\  & 1.98524\   &\ 0.76533\  \\
\ $d_2$\ &\ 0.45614\ & 2.94311\  & 1.72105\   &\ 0.59289\  \\
\ $d_3$\ &\ 3.32682\ & 3.12550\  & 1.64902\   & ---\ \ \ \ \\
\hline
\end{tabular}
\vspace{0mm}
\end{center}

\begin{center}
\includegraphics[width=8cm]{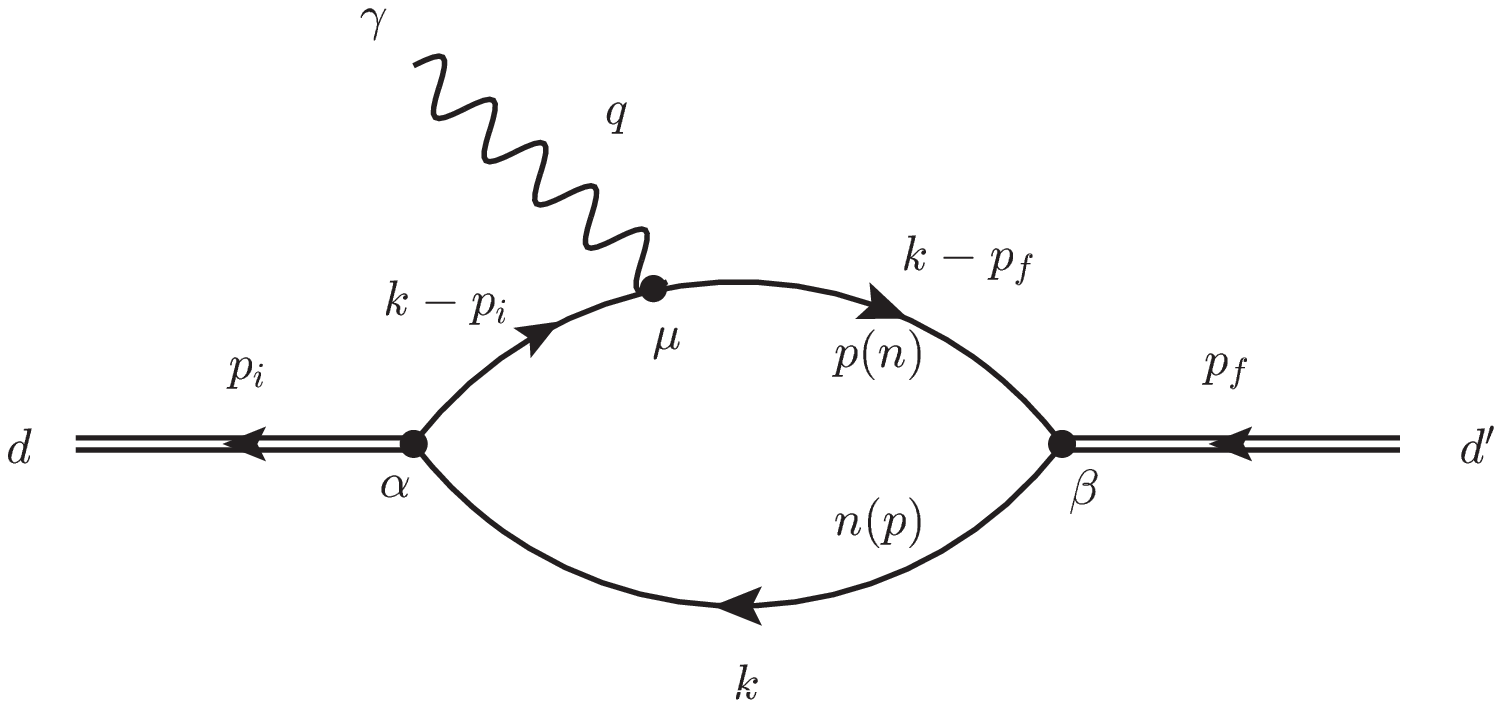}
\figcaption{\label{figure:PhotonDeuteron}  {\small Photon coupling to the deuteron.} }
\end{center}

In the Cartesian instant-form spin basis with the convention of Fig.~\ref{figure:PhotonDeuteron}, the electromagnetic current of the deuteron takes the form
\end{multicols}
\ruleup
\begin{eqnarray}
J^+_{jk}&=&i\int \frac{d^4k}{(2\pi)^4}\frac{Tr\Big [{\epsilon}^{'*\alpha}_j  {\Gamma}^{d-pn}_{\alpha}\big (k,k-p_f  \big)\big (\slashed{k}-\slashed{p}_f+m  \big) {\Gamma}^+_{\gamma-NN} (\slashed{k}-\slashed{p}_i+m  \big) {\epsilon}^{\beta}_k {\Gamma}^{d-pn}_{\beta}\big (k,k-p_i  \big) (\slashed{k}+m  \big)  \Big ]}{\big [\big( k-p_i \big)^2-m^2+i\epsilon  \big ] \big [ k^2-m^2+i\epsilon  \big ] \big [\big( k-p_f \big)^2-m^2+i\epsilon  \big ]}
\nonumber\\
&&\times
\Lambda(k,p_f)\Lambda(k,p_i)\Lambda_0(\lambda^2,Q^2) \ .
\label{eq:LoopCurrent}
\end{eqnarray}
\ruledown \vspace{0.5cm}
\begin{multicols}{2}
\noindent
It should be reiterated that the polarization vectors of $\epsilon^\beta_k$ and $\epsilon^{'\alpha}_j$, defined in Eqs.~(\ref{eq:intpv}) and (\ref{eq:finpv}), are in the Cartesian instant-form spin basis.

The regularization function in the above Eq.~(\ref{eq:LoopCurrent}) is taken as the sum of the two dipoles
\begin{eqnarray}
\Lambda(k,p)&=&N_1\bigg (\frac{1}{[(k-p)^2-m^2_R+i\epsilon]^2}
\nonumber\\
&&
-\frac{N'}{[(k-p)^2-m^2_{R2}+i\epsilon]^2}\bigg ) \ ,
\end{eqnarray}
where $m_R$ and $m_{R2}$ are the two independent regulator masses; $N'$ is a model-dependent parameter and the normalization constant $N_1$ can be obtained by $G_0(0)=1$. Here we use the two monopole functions instead
of one ~\citep{Frederico1997} due to the fitting to the deuteron experimental data.

The additional regularization function is
\begin{eqnarray}
\Lambda_0(\lambda^2,Q^2)&=&\frac{\lambda^2}{\lambda^2+Q^2},
\label{eq:lambda0}
\end{eqnarray}	
with $\lambda^2$ being a model-dependent parameter. This regularization function is also needed in order to suppress the $Q^2-$dependences of the form factors in the larger $Q^2$ region.

Let's look at the poles of the integral. In the Breit frame, $p_i^+ = p_f^+= p^0\equiv p^+$, and for the condition of $p^+>k^+>0$, only one pole contributes to the final residue of Eq.~(\ref{eq:LoopCurrent}),  that is
\begin{eqnarray}
k^-=\frac{{\bf k}^2_{\perp}+m^2-\imath\epsilon}{k^+}\equiv \overline{k}\ .
\label{eq:pole}
\end{eqnarray}
Detailed discussion of the poles and residues of the integral is referred to Refs.~\cite{Frederico1997, Liang2015,Miller2009}.

Similar to Refs.~\citep{Pereira2007} and \citep{Frederico1997}, after carrying out the integration of $k^-$ and $x$, we can rewrite the propagators together with the corresponding regulators $\Lambda(k,p)$ as the light-front wave functions. The initial light-front wave function is obtained as
\end{multicols}
\ruleup
\begin{eqnarray}
\frac{1}{(k-p_i)^2 - m^2+\imath\epsilon}
\bigg (\frac{1}{[(k-p_i)^2 - m^2_R+\imath\epsilon]^2}
-\frac{1}{[(k-p_i)^2 - m^2_{R2}+\imath\epsilon]^2 }\bigg )
\nonumber\\
=\frac{1}{(1-x)^3(m^2_D-M_0^2)}
\bigg( \frac{1}{(m^2_D- M^2_R)^2}-\frac{1}{(m^2_D- M^2_{R2})^2}\bigg )
\equiv \textit{$\phi$}_i(x,{\bf k}_\perp)\ ,
\label{eq:den}
\end{eqnarray}
\ruledown \vspace{0.5cm}
\begin{multicols}{2}
\noindent
where the mass squared $M_0^2$ is given by
\begin{eqnarray}
M^2_0= \frac{{\bf k}^2_\perp+m^2}{x}
+\frac{({\bf p_i}-{\bf k})^2_\perp+m^2}{1-x}-{\bf p_i}_\perp^2,
\end{eqnarray}
with ${\bf p_i}_\perp=(p_i^1,p_i^2)$, and the functions $M_R^2$ and $M_{R2}^2$ are
\begin{eqnarray}
M^2_R&=& \frac{{\bf k}^2_\perp+m^2}{x}
+\frac{({\bf p_i}-{\bf k})^2_\perp+m^2_R}{1-x}-{\bf p_i}_\perp^2 \ , \
\nonumber\\
M^2_{R2}&=& \frac{{\bf k}^2_\perp+m^2}{x}
+\frac{({\bf p_i}-{\bf k})^2_\perp+m^2_{R2}}{1-x}-{\bf p_i}_\perp^2 \ .
\end{eqnarray}
To get the corresponding final light-front wave function $\textit{$\phi$}_f$, one only needs to do the replacement of $p_i \leftrightarrow p_f$. In terms of the initial and final light-front wave functions,
the current $J^+_{jk}$ has the form
\begin{eqnarray}
J^+_{jk}&=&i\int \frac{d^2{\bf k}_\perp dx}{(2\pi)^4}{\it \mathcal{N}}^+_{jk} \textit{$\phi$}^*_f(x,{\bf k}_\perp)\textit{$\phi$}_i(x,{\bf k}_\perp)
\Lambda_0(\lambda^2,Q^2) \ , \nonumber\\
\label{eq:LoopCurrentPhi}
\end{eqnarray}
where $\it \mathcal{N}^+_{jk}=Tr[{\epsilon}^{'*\alpha}_j  {\Gamma}^{d-pn}_{\alpha}\big (k,k-p_f  \big)\big (\slashed{k}-\slashed{p}_f+m  \big) {\Gamma}^+_{\gamma-NN} (\slashed{k}-\slashed{p}_i+m  \big) {\epsilon}^{\beta}_k {\Gamma}^{d-pn}_{\beta}\big (k,k-p_i  \big) (\slashed{k}+m \big)]\arrowvert_{k^- = \overline{k}}$. The above
light-front wave functions correspond to the wave function of the $S$ wave state~\citep{Jaus1990}.

\section{Numerical results and discussions}

So far, in our calculation there are 5 model-dependent parameters, i.e., two regulator masses $m_R$ and $m_{R2}$, a normalization constant $N'$, a regulator constant $\lambda^2$ and $\rho$. Moreover, the requirement of stability of the bound states, mentioned in the work of Ref.~\cite{Frederico1997}, should also be maintained. This constrains  $m+m_{R2}>m_D$ and $m+m_{R}>m_D$. By fitting to the experimental data of deuteron form factors from Ref.~\citep{Abbott2000} and its references, we take the parameter values shown in Table~\ref{tab:parameters}. The parameter errors are obtained through the propagation of experimental data  errors, under the linear approximation. Usually, to get the most appropriate parameters, the initial values must be chosen to be as close as possible. In this work, we are trying to describe all three form factors and three static properties
simultaneously with only five parameters, which are not easy to fit equally well. Besides, only a few restricted conditions can help to narrow down the parameter space. Therefore, the small errors in the parameters, all less than 1$\%$, may just mean that this set of values is very close to the optimal one. Out of the range, the calculation results would deviate from the experimental data quickly since the integrals are sensitive to the model parameters. The three form factors $G_0$, $G_1$ and $G_2$ are shown in Figs.~\ref{fig:GcGmGqSD0011L04-G0}, ~\ref{fig:GcGmGqSD0011L04-G1}, and ~\ref{fig:GcGmGqSD0011L04-G2} as the functions of $Q^2$, together with experimental data from Ref.~\citep{Abbott2000,The1991,Nikolenko2003,Karpius2005,Zhang2011,Martin1977,Platchkov1990}.

\begin{center}
\tabcaption{ \label{tab:parameters} {\small The parameters used in this work.}}
\footnotesize
\begin{tabular*}{60mm}{c|c}
\toprule Parameter & Value \\ \hline
$m_R$\       &\ $2.238\pm0.010~$GeV\ \\
$m_{R2}$\    &\ $1.251\pm0.004~$GeV\ \\
N'\          &\ $0.135\pm0.004$\ \ \ \ \ \ \ \ \\
$\rho$\      &\ $-0.011\pm1.33$e-6\ \ \ \ \ \ \\
$\lambda^2$\ &\ $0.40\pm0.012$GeV$^2$\ \\
\bottomrule
\end{tabular*}
\vspace{0mm}
\end{center}
\vspace{0mm}

The experimental data~\citep{Garcona2001} shows that the value of the deuteron
quadrupole moment $Q_d(=0.286 $fm$^2)$ is smaller than the mean square charge radius $<r^2>(=4.54 $fm$^2)$ by more than one order of magnitude. This feature implies that the ground state of the deuteron is basically spherically symmetric and the admixture of the $D$ wave accounts for a relatively small part~\citep{Gourdin1974}. According to the results of most three-dimension potential models, the $D$ wave probability ranges from 4.83\% to 5.8\%~\citep{Garcona2001}. In this work, $\rho^2$ plays a similar but not identical role. The meaning of $\rho$, here, is not exactly the same as  in the three-dimensional potential models, and it may also be different from our previous calculation which was done in Euclidean space with a Gauss-type regularization~\citep{Dong2008,Liang2015}. It is the value of $\rho^2$ that actually connects to the $D$ wave admixture. Therefore, the sign of the $\rho$, in principle, only affects the interference terms between $S$ and $D$ waves,
and the negative sign is allowed.

In Figs.~\ref{fig:GcGmGqSD0011L04-G0}$\sim$\ref{fig:GcGmGqSD0011L04-G2}, the curves represent our calculated results.  Figure~\ref{fig:GcGmGqSD0011L04-G0} shows that the charge form factor $G_0$ obtained in our light-front calculation does not provide the depth at $Q^2\approx0.7$ GeV$^2$.
Moreover, Fig.~\ref{fig:GcGmGqSD0011L04-G1} indicates that our  light-front calculation for the  magnetic form factor $G_1$ fits the experimental data very well at $Q^2=0\sim0.6 $ GeV$^2$. For the $Q^2 > 0.6 $ GeV$^2$ region, the obtained results became
larger than the experimental data. Finally, Fig.~\ref{fig:GcGmGqSD0011L04-G2} tells that the calculated quadrupole form factor $G_2$ is somewhat lower than the experimental data. But overall, we conclude that our simulations of $G_1$ and $G_2$ are reasonable. In our calculation, we also test all the four different prescriptions for the form factor extraction, and we find that, in the small momentum transfer region, the differences among the four prescriptions for the deuteron form factors are negligible. This conclusion is unlike the case of the $\rho$ meson in Ref.~\cite{Frederico1997}. It coincides with the work in Ref~\citep{Frankfurt1993}. Here, in Figs.~\ref{fig:GcGmGqSD0011L04-G0}$\sim$\ref{fig:GcGmGqSD0011L04-G2},
we only plot our calculated results with the FFS prescription.

We also calculate the deuteron static properties by using the relations of Eq.~(\ref{eq:LowEO}) with the same set of parameters mentioned above. The calculated results are showed in Table~\ref{tab:observables}. Similar to
the three form factors with the four prescriptions,  the differences among the four prescriptions for the three low-energy observables are also negligible. We also find that in the light-front calculation the obtained deuteron charge radius $<r>$, magnetic moment $\mu_d$, and the quadrupole moment $Q_d$ are about $15\% \sim 20\%$ larger than the experimental values.

\begin{center}
\includegraphics[width=8cm]{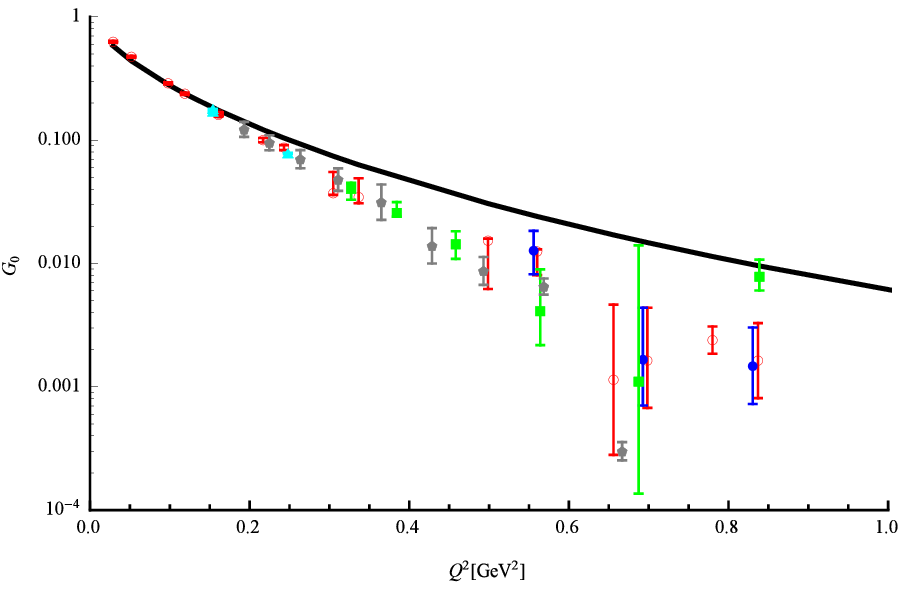}
\figcaption{\label{fig:GcGmGqSD0011L04-G0}{\small The obtained deuteron charge form factor $G_0$. The points with error bars are the experimental data from Ref.~\cite{Abbott2000}(red circle),~\cite{The1991}(blue disk),~\cite{Nikolenko2003}(green rectangle),~\cite{Karpius2005}(cyan, triangle) and~\cite{Zhang2011}(gray pentagon). The curve shows our results.}}
\end{center}

\begin{center}
\includegraphics[width=8cm]{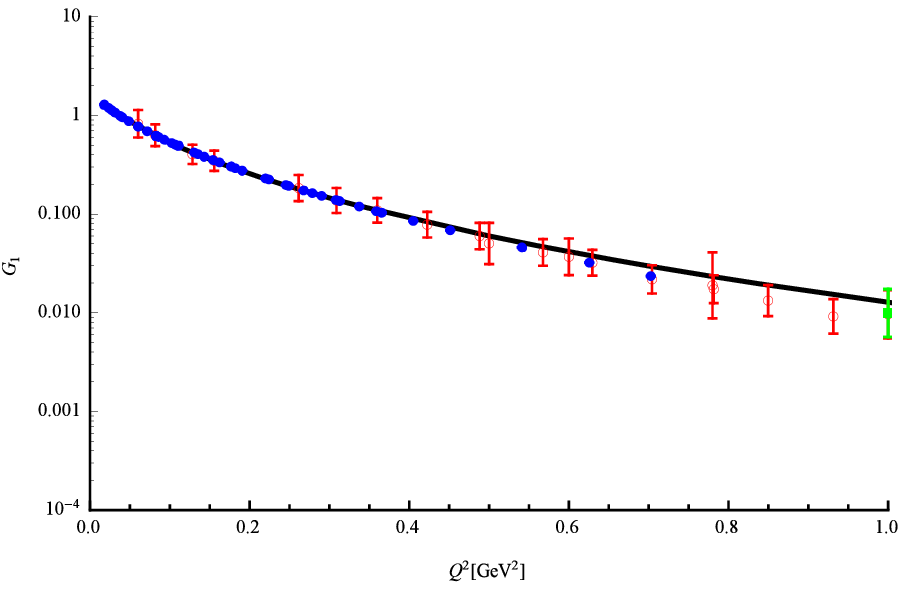}
\figcaption{\label{fig:GcGmGqSD0011L04-G1}{\small The obtained deuteron magnetic form factor $G_1$. The notations are
the same as Fig.~\ref{fig:GcGmGqSD0011L04-G0}. The data are from Ref.~\cite{Abbott2000}(red circle), ~\cite{Platchkov1990}(blue disk), ~\cite{Martin1977}(green rectangle).}}
\end{center}

\begin{center}
\includegraphics[width=8cm]{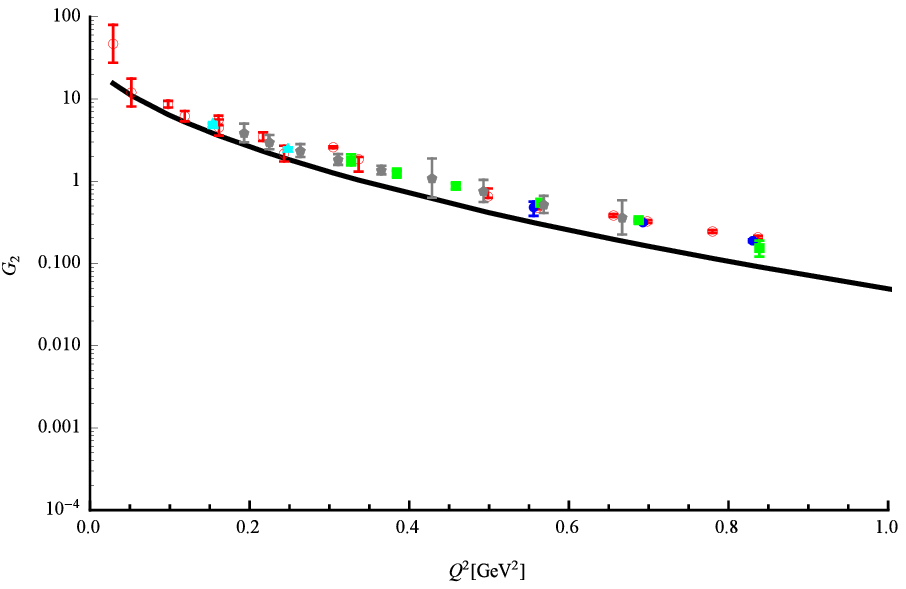}
\figcaption{\label{fig:GcGmGqSD0011L04-G2}{\small The obtained deuteron quadrupole form factor $G_2$. The notations are the same as Fig.~\ref{fig:GcGmGqSD0011L04-G0}.
The data are from Ref.~\cite{Abbott2000,The1991,Nikolenko2003,Karpius2005,Zhang2011}, marked the same as in Fig.~\ref{fig:GcGmGqSD0011L04-G0}.} }
\end{center}

\begin{center}
\tabcaption{ \label{tab:observables} {\small The calculated deuteron static properties.}}
\footnotesize
\begin{tabular*}{85mm}{c@{\extracolsep{\fill}}ccccc}
\toprule Model & FFS & GK & CCKP & BH & Experiment~\citep{Garcona2001} \\
\hline
$<r> ($fm$)$ & 2.58 & 2.58 & 2.58 & 2.58 & 2.130(10) \\
$\mu_d/{\mu}_N$ & 1.024 & 1.024 & 1.024 & 1.024 & 0.8574382284 (94) \\
$Q_d ($fm$^2)$ & 0.325 & 0.325 & 0.325 & 0.325 & 0.2859 (3) \\
\bottomrule
\end{tabular*}
\vspace{0mm}
\end{center}
\vspace{0mm}

From our numerical calculation, we find that the pure $S$ wave vertex plays important roles in the charge $G_0$ and the magnetic $G_1$ form factors.  In the small momentum transfer region, the contribution accounts for about 65\% and 75\% respectively of the total of $G_0$ and $G_1$  where both the $S$ and $D$ wave vertexes are taken into account.  This feature also implies that the $D$ wave vertex also contributes to the $G_0$ and $G_1$ form factors. Because of the above analyses, we perform our fitting to the data of all three form factors by taking both the $S$ and $D$ wave vertexes simultaneously, and thus we get the optimum values of the model-dependent parameters as mentioned above. In addition, in contrast with the first term of $\Gamma^{\mu}_S$(i.e. $\gamma^{\mu}$), we find that the contribution of the second term to $G_0$ and $G_1$  can be neglected.

Figure~\ref{fig:Swave2part-G2} shows that the pure $S$ wave vertex only contributes a small part to the  quadrupole $G_2$ form factor,  and there exists strong cancellation between the two terms in $S$ wave vertex function. The $D$ wave vertex together with the interference terms between  the $D$ wave and $S$ wave vertexes, on the other hand, account much more for the $G_2$ form factor. This conclusion is consistent with the potential model calculation~\citep{Machleidt2001} and agrees with the general knowledge that the quadrupole moment originates from the non-central force between the two nucleons~\citep{Garcona2001}.

Actually, the second term in $\Gamma^{\mu}_D$ can be rewritten as
\begin{eqnarray}
\Gamma^\mu_{D2}&\equiv&-\frac{3\rho}{m\delta}\left(-\left(\slashed{k}- \frac{\slashed{p}_{i,f}}{2} \right){\gamma}^{\mu}+2\left(k- \frac{p_{i,f}}{2}\right)^\mu\right)\nonumber\\
&\equiv&-\frac{3\rho}{m\delta}\left(\Gamma^\mu_{D21}+\Gamma^\mu_{D22}\right).
\end{eqnarray}
One may eliminate $\Gamma^\mu_{D21}$ term in both initial and final vertex functions in order to distinguish the contributions of these two terms to $G_2$. The relevant results are plotted  in Fig.~\ref{fig:SD0011L04d22-G2}. Comparing the calculated results showed in Figs.~\ref{fig:Swave2part-G2} and \ref{fig:SD0011L04d22-G2}, we find that the contribution of $\Gamma^\mu_{D22}$ is essential to the quadrupole $G_2$ form factor, which contributes over 84\% of the integral. Namely, it is the structure which proportional to $k^\mu$ in the $D$ wave function that mainly accounts for the quadrupole form factor.

\begin{center}
\includegraphics[width=8cm]{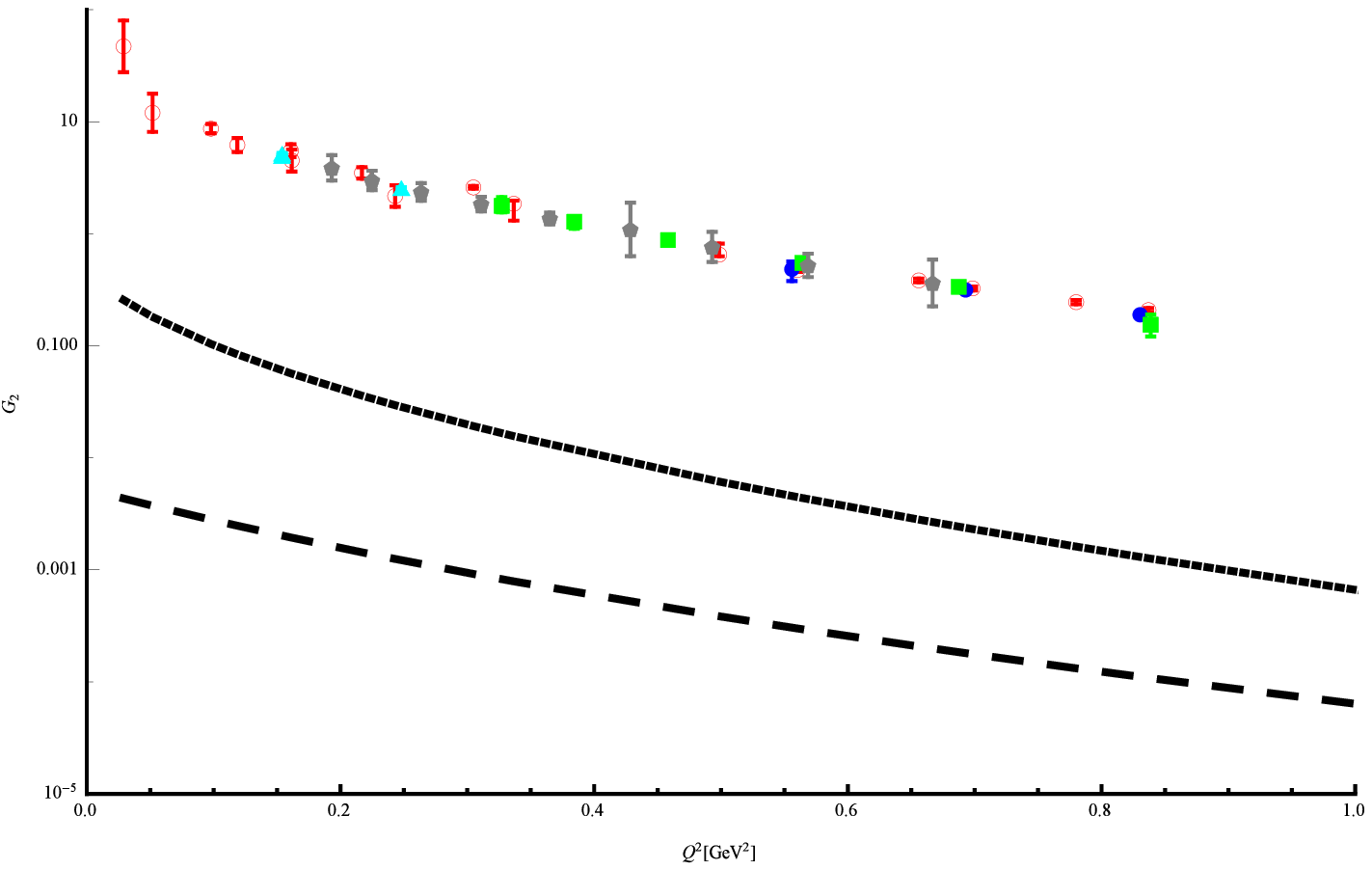}
\figcaption{\label{fig:Swave2part-G2} {\small The obtained deuteron quadrupole form factor $G_2$ from $S$ wave vertex only. The dashed curve is the contribution from the full $S$ wave vertex and the dotted curve is obtained by leaving only $\gamma^\mu$ terms in both the initial and final vertex functions.}}
\end{center}

\begin{center}
\includegraphics[width=8cm]{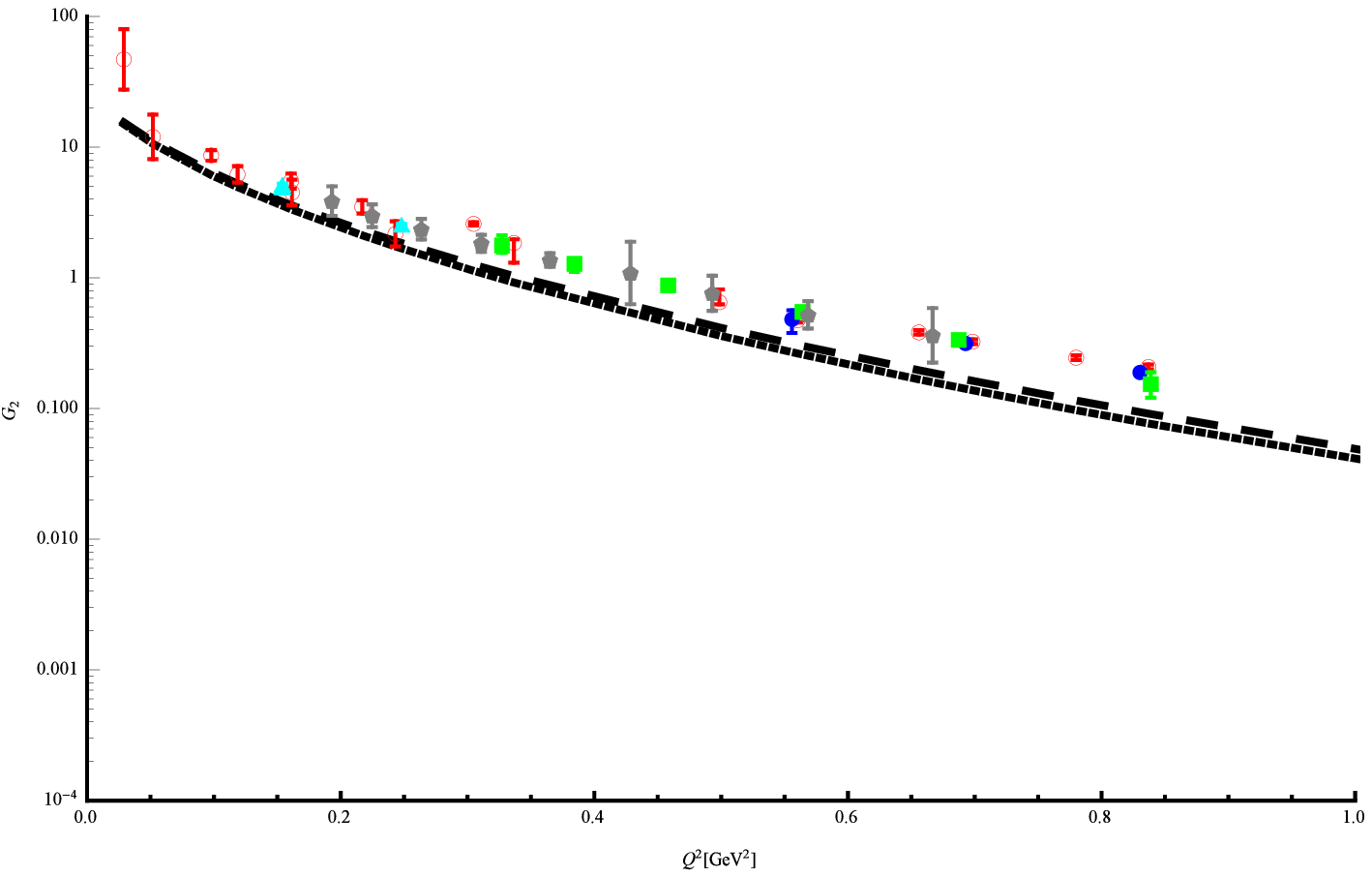}
\figcaption{\label{fig:SD0011L04d22-G2} {\small The obtained deuteron quadrupole form factors $G_2$.
The dotted curve  shows the results without the $\Gamma^\mu_{D21}$ term in both initial and final vertex functions, and the dashed curve is for the full $S$ and $D$ wave vertexes.} }
\end{center}

\section{Summary}

In this work, the electromagnetic form factors and other low-energy observables of the  deuteron are studied with the help of the light-front approach, where the deuteron is regarded as a weakly bound state of a proton and a neutron. We take into account both the $S$ and $D$ waves interacting vertexes among the deuteron, proton and neutron, by introducing phenomenological vertex functions. Here we intend to simulate the momentum distribution inside the deuteron by these vertex functions and the regularization functions. The parameters are obtained by fitting to the experimental data of the three form factors with the FFS extraction prescription. We compare our calculated results among the four prescriptions of the form factors extraction, and find, in the small momentum transfer region, the differences among the four are negligible. We also calculate other low-energy observables for the deuteron and we see negligible differences among the four prescriptions. In our calculations, the $S$ wave vertex function is assumed to be the same as $\rho-q\overline{q}$ in Ref.~\citep{Frederico1997}. The
contribution of the $D$ wave vertex function, which results from the tensor force, is studied in detail and we conclude that it mainly corresponds to the quadrupole form factor. Our numerical results show that the light-front approach can only roughly reproduce the deuteron electromagnetic form factors, like $G_0$, $G_1$ and $G_2$, in the
low $Q^2$ region. Moreover, the estimated low-energy observables for the deuteron are all overestimated by about $15\%\sim 20\%$ compared to the data.

Although the light-front approach is considered suitable for describing  bound state problems, the present results for the deuteron are not very satisfactory. This is due to the complicated structure of the deuteron. It should be mentioned that Ref.~\citep{Gustafsson2006} develops a phenomenological parametrization model to get a good fitting to the deuteron form factors. In their fitting, different sets of the parameters are employed for the different form factors. We expect to further improve our theoretical calculations by adjusting the structures of the vertex functions and the sophisticated regularization functions. In addition, other ingredients, like the two-body current (see Ref.~\citep{Dong2008}), can also be taken into account.

\end{multicols}

\vspace{-1mm}
\centerline{\rule{80mm}{0.1pt}}
\vspace{2mm}

\begin{multicols}{2}

\end{multicols}

\clearpage
\end{CJK*}
\end{document}